\documentclass[aps,prl,superscriptaddress,preprintnumbers,twocolumn,groupedaddress,showpacs]{revtex4}
\usepackage{graphicx}

\def\beq{\begin{equation}}
\def\eeq{\end{equation}}
\def\bea{\begin{eqnarray}}
\def\eea{\end{eqnarray}}

\begin{document}

\title{  QCD NLO and EW NLO corrections to $t\bar{t}H$ production with top quark decays at hadron collider  }
\author{ Zhang Yu, Ma Wen-Gan, Zhang Ren-You, Chen Chong, and Guo Lei \\
{\small  Department of Modern Physics, University of Science and Technology of China,} \\
{\small  Hefei, Anhui 230026, P.R.China}}

\begin{abstract}
Higgs boson production associated with a top quark pair is an important process in studying the nature of the newly discovered Higgs boson at the LHC. In this letter, we report on our calculations including the next-to-leading order (NLO) QCD and NLO electroweak corrections to the $pp\to t\bar{t}H$ process in the standard model. We present the integrated cross sections at the $14~{\rm TeV}$ LHC and even at the future proton-proton colliders with $\sqrt{s}=33$ and $100~{\rm TeV}$. Our calculation includes the top quark subsequent decays by adopting the narrow width approximation. The kinematic distributions of Higgs boson and top quark decay products at the LHC are provided. We find that the ${\cal O}(\alpha_s^2\alpha_{ew}^2)$ corrections are quantitatively comparable with the ${\cal O}(\alpha_s^3\alpha_{ew})$ corrections in some kinematic region.
\end{abstract}


\maketitle

\par
{\it Introduction.-}
\label{Sec:intro}
In July 2012 both the ATLAS and CMS collaborations at the LHC reported the evidence of the existence of a new neutral boson with mass of around $126~{\rm GeV}$ \cite{atlas,cms} in searching for the standard model (SM) Higgs boson \cite{higgs1,higgs2,higgs3}. They provided very clear evidence to strengthen the hypothesis that the newly discovered particle is the SM Higgs boson \cite{smh1,smh2}. To understand the nature of the discovered Higgs boson is one of the major goals of the LHC, especially its Yukawa coupling to the heaviest fermion, top quark ($y_t$). Unlike the dominant production mechanism for the Higgs boson at the LHC, loop production mechanism $gg\to H$,  which is sensitive to $y_t$ but can be easily polluted by the particles beyond the SM running in the loop, the $t\bar{t}H$ associated production can be used to probe the structure of top-Higgs interactions unequivocally.

\par
From the experimental point of view, the analysis of the $t\bar{t}H$ production event is extremely challenging. One of the difficulties is related to the production rate which is strongly suppressed by parton distribution functions (PDFs) due to the production of three heavy particles needs a large center-of-mass collision energy for the initial partons. Other difficulties are manifested by the presence of various irrepressible backgrounds and by the complexity of the final state, which make its kinematic reconstruction far from straightforward. Current search strategies are mainly designed for the $H \to b\bar{b}$ decay mode combining with the fully leptonic and/or semi-leptonic decay channels for the top quark \cite{hbb1,hbb2,hbb3}. The other two important Higgs boson decay channels $H\to WW^*$ \cite{hww1,hww2,hww3} and $H \to \tau \tau$ \cite{htau1,htau2,htau3,htau4} have also been discussed. The process $pp \to t\bar{t}H$ has preliminarily been searched by the ATLAS \cite{searcha1,searcha2} and CMS \cite{searchc1,searchc2} using data samples collected at the LHC with the center-of-mass energies of $7~{\rm TeV}$ and $8~{\rm TeV}$, but the current luminosity and analyses have not reached the sensitivity required by the SM Higgs boson.

\par
In completely determining the nature of the Higgs boson at the LHC, precision theoretical predictions are necessary and will play a crucial role. The leading order (LO) predictions for $t\bar{t}H$ production at ${\cal O}(\alpha_s^2\alpha_{ew})$ have been given some years ago \cite{lo}. The Higgs production in association with a top pair was studied up to QCD next-to-leading order (NLO) accuracy at ${\cal O}(\alpha_s^3\alpha_{ew})$ in Refs.\cite{nlo1,nlo2,nlo3,nlo4}. The predictions for the $t\bar{t}H$ production with parton shower and hadronization effects at the LHC are provided in Refs.\cite{ps1,ps2}. Further discussions on the uncertainties of the scale, $\alpha_s$, and PDF are included in Refs.\cite{higgs-handbook1,higgs-handbook2,higgs-handbook3}. Besides, the complete NLO QCD corrections to the process $pp\to t\bar{t} H + 1 jet$ are also given in Ref.\cite{tthj}.

\par
In order to meet requirement of the experimental measurement, the accuracy up to QCD NLO plus electroweak (EW) NLO for the $pp \to t\bar{t}H$ process with top decays is necessary which is desired in the Les Houches NLO wishlist \cite{wishlist}. Although the NLO EW correction is normally suppressed by the smallness of the coupling constant $\alpha_{ew}$ and nominally subdominant with respect to the QCD contributions, the NLO EW correction can become significant in the high-energy domain due to the appearance of Sudakov logarithms that result from the virtual exchange of soft or collinear massive weak gauge bosons \cite{sudakov, fadin, ciafaloni,pozzorini}. In this letter we calculate the NLO QCD and NLO EW corrections to the $pp \to t\bar{t}H$ process in the SM. We provide the integrated cross sections at the $14~{\rm TeV}$ LHC and at the future proton-proton colliders with $\sqrt{s}=33$ and $100~{\rm TeV}$, and study some kinematic distributions of final particles after top quark subsequential decays by adopting the narrow width approximation (NWA).

\par
{\it Calculational setup.-}
\label{Sec:setup}
Since the calculation strategy of the NLO QCD correction to $pp \to t\bar{t}H$ process has been already provided in Refs.\cite{nlo1,nlo2,nlo3,nlo4}, here we give only the calculation setup of the NLO EW correction. Our calculation for the $pp \to t\bar{t} H$ process is carried out in the 't Hooft-Feynman gauge. We adopt the dimensional regularization scheme in the NLO calculations, where the dimensions of spinor and space-time manifolds are extended to $D=4-2\epsilon$ to isolate the ultraviolet (UV) and infrared (IR) singularities. We apply FeynArts-3.7 package \cite{feynarts} to automatically generate the Feynman diagrams and the FormCalc-7.2 program \cite{formcalc} to algebraically simplify the corresponding amplitudes. In order to solve the serious unstable problem in the calculation of the scaler and tensor integrals, we have to adopt quadruple precision arithmetic which would consume much more computer CPU time. We modified the LoopTools-2.8 package \cite{formcalc,cc} by adopting the segmentation method analogous to that in Refs.\cite{unstable-problem-1, unstable-problem-2} to treat the unstable problem and improve the efficiency in the numerical calculation of the scaler and tensor integrals. With this method the program can automatically switch to the quadruple precision codes in the region of $\frac{{\rm det} G_3}{(2 k_{max}^2)^3} < 10^{-5}$, where ${\rm det} G_3$ is the Gram determinant and $k_{max}^2$ the maximum of the external four-momentum squared for a given 4-point integral.

\par
The value of fine structure constant is obtained by adopting the $G_{\mu}$-scheme via $\alpha_{ew}=\frac{\sqrt{2} G_{\mu} M^2_W}{\pi} \left(1-\frac{M^2_W}{M^2_Z} \right)$. Compared to the $\alpha_{ew}$-scheme defined in the Thomson limit ($Q=0$), choosing the $G_{\mu}$-scheme can avoid large logarithms of the light fermion masses generated by running of the coupling constant $\alpha_{ew}(Q)$ from the scale $Q=0$ to the EW scale $Q=M_W$ in the NLO corrections. In addition, the counterterm of coupling constant $\alpha_{ew}$ in $G_{\mu}$-scheme inherits a correction term $\Delta r$ from the weak corrections to muon decay \cite{deltar}. The relevant fields and masses are renormalized by adopting the on-mass-shell renormalization scheme and the explicit expressions for the renormalization constants are detailed in Ref.\cite{denner}.

\par
The parent process $p p \to t\bar{t} H$ is contributed by $gg \to t\bar{t} H$ and $q \bar q \to t\bar{t} H$ partonic processes, and the NLO EW correction contains the virtual and real emission correction components. The ultraviolet (UV) divergences in loop integrals for both partonic processes are regularised dimensionally. After performing the renormalization procedure, the whole NLO EW correction is UV finite.

\par
For the NLO EW correction to the subprocess $gg \to t\bar{t} H$, the photonic IR divergences originating from exchange of virtual photon in loop can be extracted and cancelled with those in the real photon emission correction by employing either the dipole subtraction (DS) method \cite{dipole1,dipole2,dipole3} or the two cutoff phase space slicing (TCPSS) method \cite{tcpss}. In our calculations, we transfer the dipole formulae in QCD provided in Ref.\cite{dipole1} in a straightforward way to the case of dimensionally regularised photon emission.

\par
Another dominant $t\bar{t}H$ production process occurs via QCD-mediated $q\bar{q}$ annihilation. In a similar way as used for previous subprocess, the photonic IR divergences in the virtual corrections to the $q\bar{q} \to t\bar{t} H$ subprocess can be compensated by those from the real emission processes $q \bar q \to t \bar t H \gamma$, $\gamma q(\bar q) \to t \bar t H q(\bar q)$ and the PDF counterterms in the NLO EW calculation.

\par
\begin{figure}[t]
\begin{center}
\includegraphics[width=8cm]{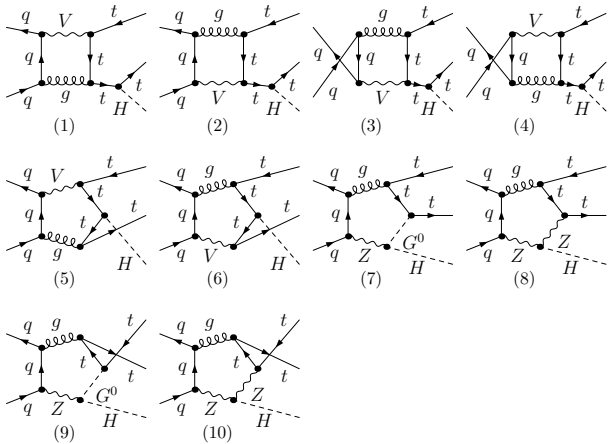}
\caption{The box (upper row) and pentagon diagrams (the lower two rows) for $q\bar{q}\to t\bar{t} H$ subprocess containing a gluon in loop, where $V=Z,\gamma$. For box diagrams, the graphs with a Higgs boson radiated off the external anti-top are not drawn. For pentagon diagrams, the graphs by exchanging the initial quark and anti-quark are not drawn.}
\label{Fig:loopg}
\end{center}
\end{figure}
\begin{figure}[t]
\begin{center}
\includegraphics[width=8cm]{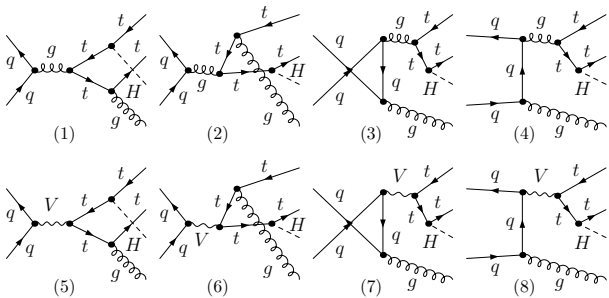}
\caption{Some representative tree-level diagrams for gluon bremsstrahlung subprocess $q\bar{q}\to t\bar{t} H+g$ at the ${\cal O}(\alpha_s^{3/2} \alpha_{ew}^{1/2})$ (upper row) and ${\cal O}(\alpha_s^{1/2} \alpha_{ew}^{3/2})$ (lower row, where $V=Z,\gamma$).}
\label{Fig:realg}
\end{center}
\end{figure}

A specific peculiarity in the NLO EW calculation for the $q \bar q$ annihilation subprocess, is that each of the ${\cal O}(\alpha_s\alpha_{ew}^{3/2})$ box and pentagon graphs shown in Fig.\ref{Fig:loopg} contains a gluon in loop and may induce additional gluonic IR divergences. To eliminate these gluonic IR divergences in the NLO EW calculation, we have to include additionally two ${\cal O}(\alpha_s^2\alpha_{ew}^2)$ correction parts which are missing in previous calculation. One part is resulted from the interference between the ${\cal O}(\alpha_s^{3/2}\alpha_{ew}^{1/2})$ Feynman diagrams (consisting of 12 gluon emission graphs, the representative graphs are shown in upper row of Fig.\ref{Fig:realg}) and the ${\cal O}(\alpha_s^{1/2}\alpha_{ew}^{3/2})$ ones (consisting of 28 gluon emission graphs, the representative graphs are shown in lower row of Fig.\ref{Fig:realg}). Here we find that only the contribution from the interference between the initial and final state gluon radiation diagrams is nonzero owing to the color structure. Another missing part at ${\cal O}(\alpha_s^2\alpha_{ew}^2)$ comes from the interference between the box (pentagon) diagrams which involve two virtual gluons in loop at ${\cal O}(\alpha_s^2\alpha_{ew}^{1/2})$, and the EW-mediated LO diagrams for $q \bar q \to t \bar t H$ subprocess at ${\cal O}(\alpha_{ew}^{3/2})$. Both the photonic IR and gluonic IR divergences are regularised dimensionally. After combining all the contributing parts at the ${\cal O}(\alpha_s^2\alpha_{ew}^2)$ mentioned above, the final result is IR finite.

\par
As we know that the LO subprocess $gq(\bar q) \to t \bar t H q(\bar q)$ can be QCD-mediated or EW-mediated by exchanging either gluon or neutral vector boson $V$ ($V=Z,\gamma$), and the contributions to its cross section contain ${\cal O}(\alpha_s^3\alpha_{ew})$, ${\cal O}(\alpha_s^2\alpha_{ew}^2)$  and ${\cal O}(\alpha_s\alpha_{ew}^3)$ parts. The first part has been already included in the real light-(anti)quark emission NLO QCD correction, the third part can be neglected due to the relative small order, while the second contribution part at ${\cal O}(\alpha_s^2\alpha_{ew}^2)$ from subprocess $gq(\bar q) \to t \bar t H q(\bar q)$ is IR finite due to the color structure, and should be involved in our NLO EW corrections.

\par
We provide the contributions to the cross section for the $pp \to t\bar{t} H$ process at ${\cal O}(\alpha_s\alpha_{ew}^2)$ from photon-gluon fusion subprocess $g\gamma \to t \bar t H $ separately.

\par
{\it Numerical results.-}
\label{Sec:result}
The total cross section at the QCD NLO has been checked with those in Refs.\cite{nlo3,nlo4}, and the coincident numerical results have been obtained by taking the same input parameters. In our numerical evaluations we take the following SM input parameters \cite{pdg}: $G_{\mu} = 1.1663787\times 10^{-5}~GeV^{-2}$, $M_{W} = 80.385~GeV$ and $M_{Z}= 91.1876~GeV$. The mass of Higgs boson is taken to be $M_H=126~GeV$, and all the quarks are massless except the top quark with $m_t=173.5~GeV$. The Cabibbo-Kobayashi-Maskawa (CKM) matrix is taken to be diagonal.

\par
We use the most recent NNPDF2.3QED PDFs in our calculations \cite{NNPDF2.3}, which consistently include QED corrections and a photon distribution function and thus allow to evaluate the contribution from the photon-induced processes. We factorize and absorb initial state gluonic (photonic) collinear singularities into the PDFs by using the $\overline{MS}~(DIS)$ factorization scheme. The value of the strong coupling constant quoted as $\alpha_s (M_Z) = 0.119$ dictated by the PDF set in five flavor scheme. The renormalization and the factorization scales are set to be equal, $\mu_R=\mu_F=m_t+\frac{1}{2}M_H$.

\par
The total NLO QCD plus NLO EW corrected integrated cross section at a hadron collider is defined as the summation of four pieces,
\bea
\label{Sigma}
\sigma_{NLO} =  \sigma^{(1)}_{LO}(\alpha_s^2\alpha_{ew})
+\Delta \sigma_{QCD}(\alpha_s^3\alpha_{ew}) \nonumber \\
 +\Delta \sigma_{EW}(\alpha_s^2\alpha_{ew}^2)+\sigma_{g\gamma}(\alpha_s\alpha_{ew}^2),
\eea
where $\Delta \sigma_{QCD}$ contains the NLO QCD corrections, $\Delta \sigma_{EW}$ is the summation of the corrections at ${\cal O}(\alpha_s^2\alpha_{ew}^2)$ described in the last section, and $\sigma_{g\gamma}$ denotes the contribution from the LO gluon-photon fusion subprocess $g\gamma\to t \bar t H$, $\sigma^{(1)}_{LO}$ and all other pieces in Eq.(\ref{Sigma}) are evaluated by using the NNPDF2.3QED NLO PDFs. We define the cross section at QCD/EW NLO as $\sigma_{QCD/EW} = \sigma^{(1)}_{LO}+\Delta \sigma_{QCD/EW}$. The corresponding relative QCD correction is given as $\delta_{QCD}=\sigma_{QCD}/\sigma_{LO}-1$, where the $\sigma_{LO}$ is the cross section at LO by adopting NNPDF2.3QED LO PDFs, and the corresponding relative genuine EW correction is defined as $\delta_{EW}=\sigma_{EW}/\sigma_{LO}^{(1)}-1$.

\par
In the EW NLO numerical calculations, we applied both the DS and TCPSS methods to isolate the IR singularities, and verified the consistence of the results from these two methods. In employing DS method, we also verified the independence on the parameter $\alpha \in (0,1]$, originally proposed in Refs.\cite{nagy1,nagy2}, which essentially controls the region of phase space over the subtracted terms, such as $\alpha=1$ means the full dipole subtraction has been considered. The formulae needed in this work have been presented in Ref.\cite{alpha}.

\par
In Table \ref{Tab:cs}, we provide the LO, NLO QCD plus NLO EW corrected integrated cross sections for $t \bar t H$ production at the $\sqrt{s}=14$, $33$ and $100~{\rm TeV}$ hadron colliders. There the corresponding relative QCD and EW corrections are also listed in the last two columns. The cross sections contributed by the $g\gamma \to t\bar{t}H$ subprocess are listed too.  We can see that the LO cross section at a hadron collider is enhanced by the NLO QCD corrections while suppressed by the NLO EW corrections, and the absolute relative NLO EW corrections are smaller than those of the NLO QCD corrections.
\begin{table}[h]
\begin{tabular}{cccccc}
\hline
\hline
$\sqrt{s}~TeV$& $\sigma_{LO}(pb)$ & $\sigma_{NLO}(pb)$  &$\sigma_{g\gamma}(pb)$& $\delta_{QCD}(\%)$ &$\delta_{EW}(\%)$\\
\hline \hline
\\ [-1.7ex]
$14$ & $0.49442(7)$   & $0.5862(23)$ &$0.00659$ &$22.6$ & $-1.03$ \\
\\ [-1.7ex]
$33$ & $3.3687(7)$    & $4.335(23)$  &$0.02930$ &$33.0$ & $-0.45$ \\
\\ [-1.7ex]
$100$  & $26.973(7)$  & $35.65(23)$  &$0.13475$ &$36.8$ & $-0.54$ \\
\\ [-1.7ex]
\hline
\hline
\end{tabular}
\caption{ The LO, NLO QCD plus NLO EW corrected integrated cross sections for $t \bar t H$ production at the $\sqrt{s}=14$, $33$ and $100~{\rm TeV}$ hadron colliders. The cross sections contributed by the subprocess $g\gamma \to t\bar{t}H$ are provided too. The relative NLO QCD and NLO EW corrections are listed in the last two columns.}
\label{Tab:cs}
\end{table}

\par
In the following, we investigate the kinematic distributions of final particles after the subsequential on-shell (anti-)top quark decays ($t \to W b \to l \nu b$ where $l=e,\mu$). In analysing the $pp \to t\bar{t}H \to W^+ b W^- \bar{b}H+X \to l^+l^-  b \bar{b} \nu\bar{\nu}H+X$ events, we use the NWA method and take the relevant branch ratios as $Br(t \to W b)=100\%$ and $Br(W \to l \nu)=10.80\%$ ($l=e,\mu$) \cite{pdg}.

\par
\begin{figure*}[htbp]
\includegraphics[width=8cm]{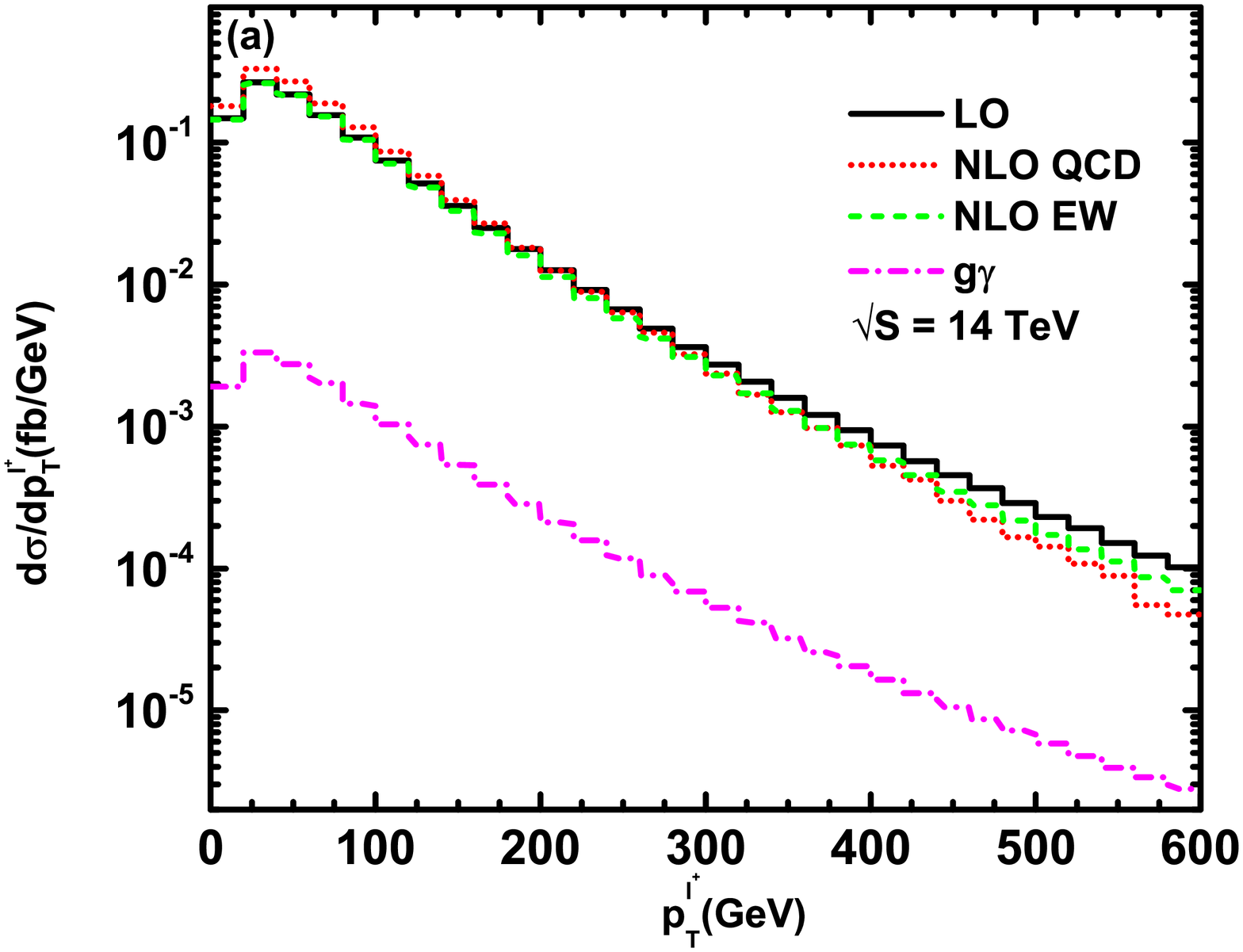}
\includegraphics[width=8cm]{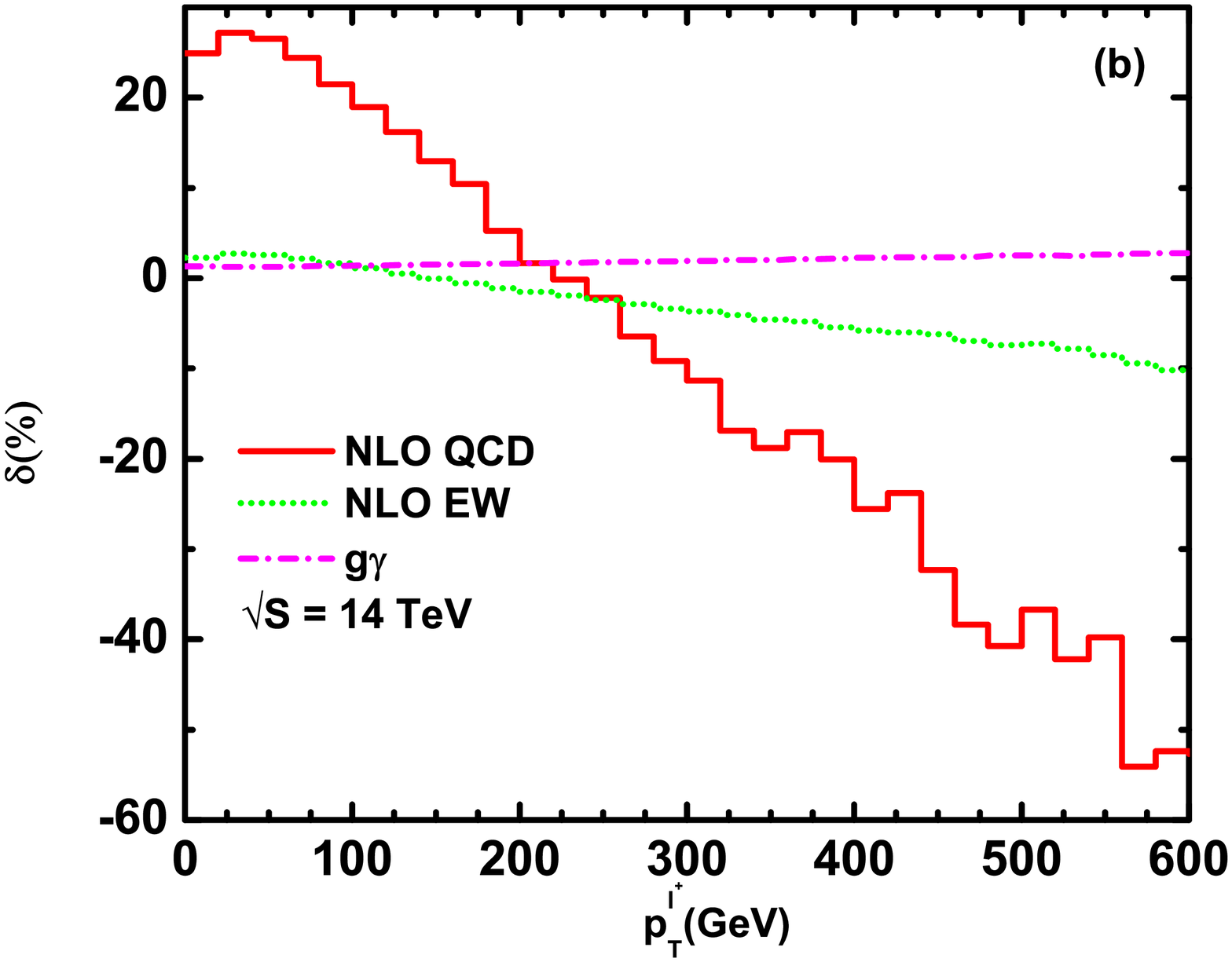}
\caption{ (a) The LO, NLO QCD, NLO EW corrected distributions of the transverse momenta of the lepton $l^+$ ($l^+ =e^+,\mu^+$) at the $14~{\rm TeV}$ LHC. The contributions from the $g\gamma \to t\bar{t}H$ subprocess are also shown there. (b) The corresponding relative corrections.  }
\label{Fig:pt}
\end{figure*}
\begin{figure*}[htbp]
\includegraphics[width=8cm]{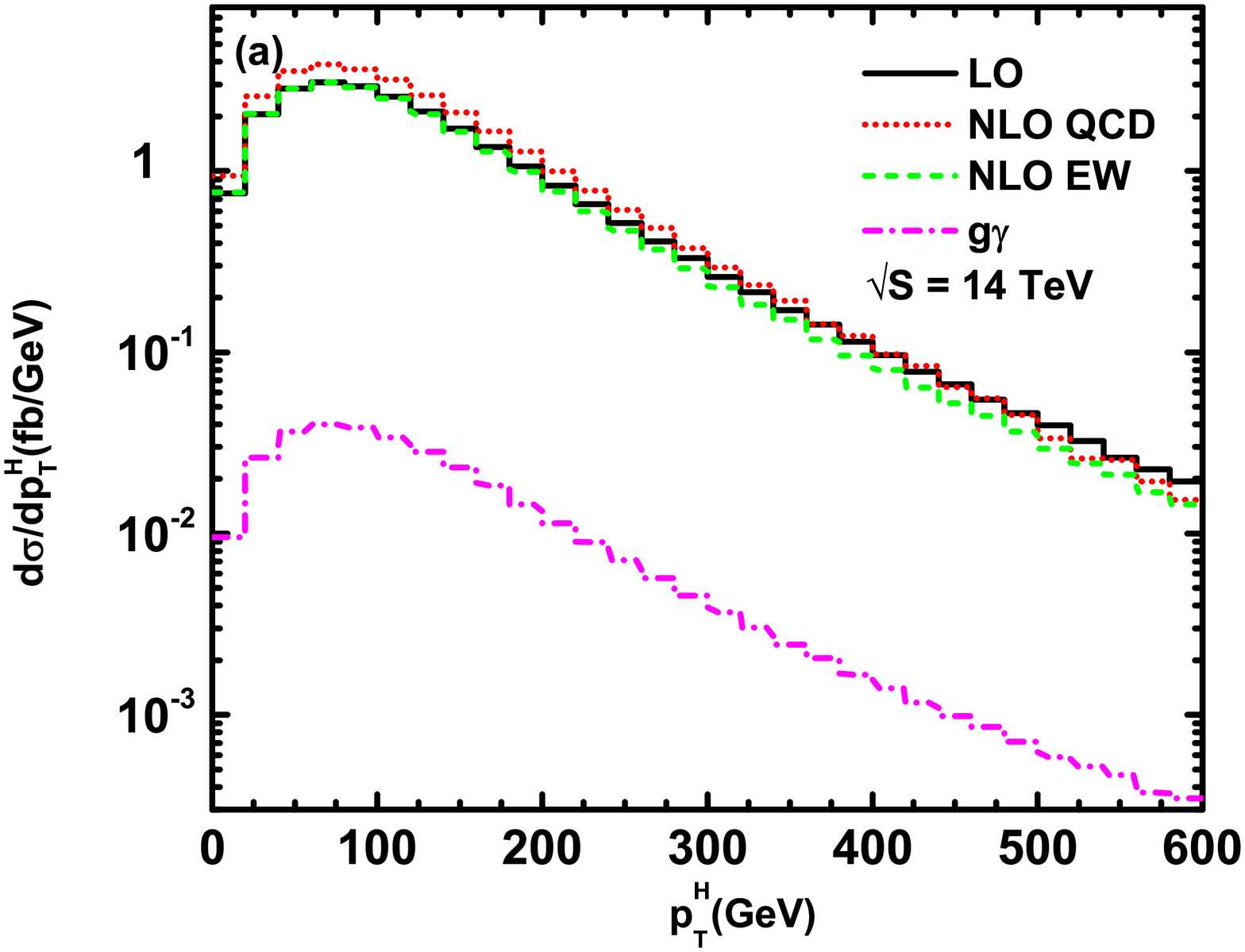}
\includegraphics[width=8cm]{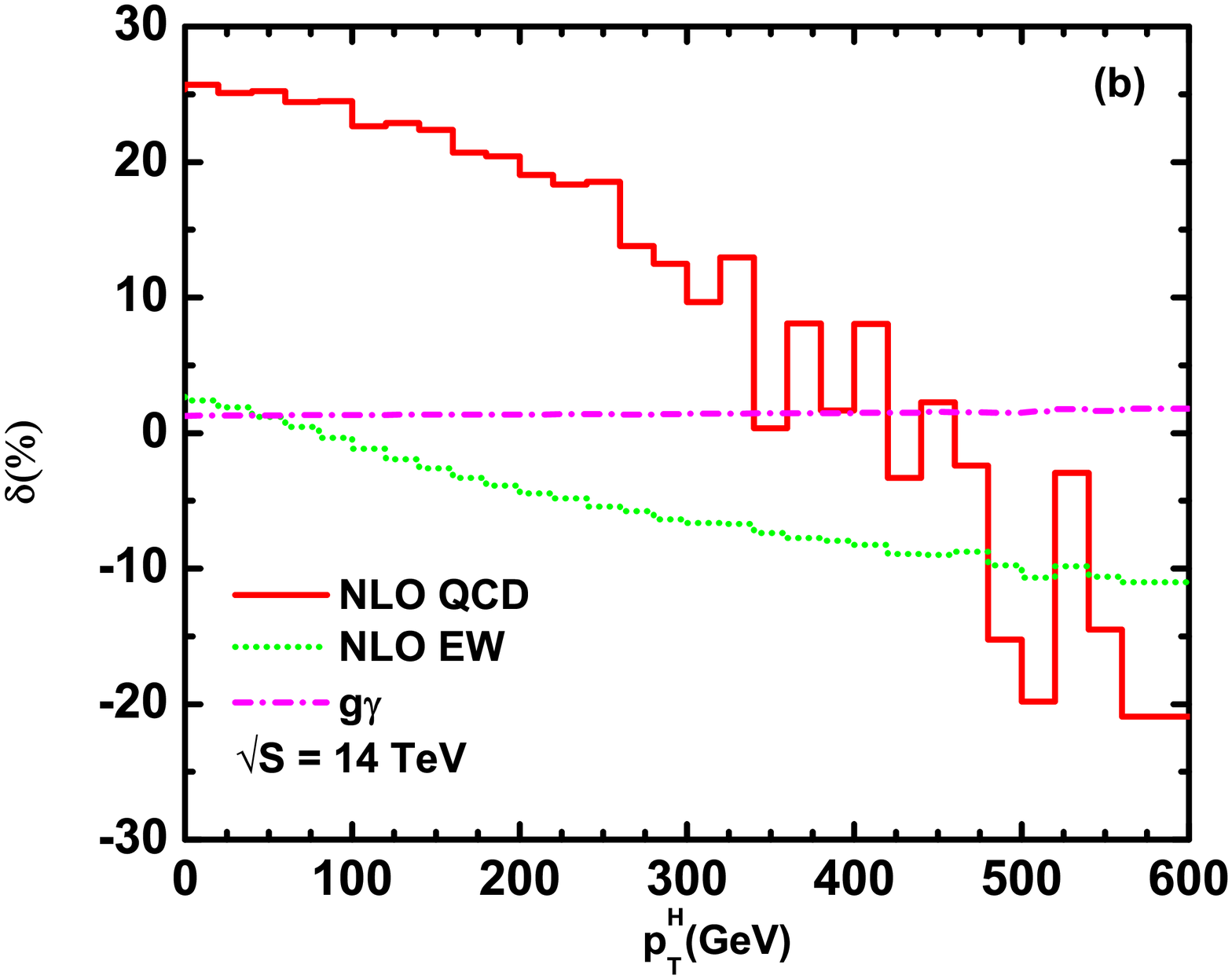}
\caption{ (a) The LO, NLO QCD, NLO EW corrected distributions of the Higgs boson transverse momenta at the $14~{\rm TeV}$ LHC.  The contributions from the $g\gamma \to t\bar{t}H$ subprocess are also shown there. (b) The corresponding relative corrections. }
\label{Fig:pt-1}
\end{figure*}

In Fig.\ref{Fig:pt}(a) and Fig.\ref{Fig:pt-1}(a), we depict the LO, NLO QCD and NLO EW corrected distributions of final positive charged lepton transverse momentum, $p_T^{l^+}$ ($l^+=e^+,\mu^+$), and Higgs boson transverse momentum, separately. Since the CP is conserved at parton level, the distribution of the $l^+$ transverse momentum should be the same as $l^-$. We can see from the figures that the differential cross sections reach maximal values at the vicinities of $p_T^{l^+} \sim 40~{\rm GeV}$ for final $l^+$ and $p_T^{H}\sim 60~{\rm GeV}$ for Higgs boson. And then the distributions drop rapidly with the increment of transverse momentum. The corresponding relative corrections are shown in Fig.\ref{Fig:pt}(b) and Fig.\ref{Fig:pt-1}(b) separately. From these figures we see that the $g\gamma$ corrections are at $1\%$ level and the relative corrections are quantitatively stable in the plotted transverse momentum ranges. Both the relative NLO QCD and EW corrections are positive at lower $p_T$ region and become negative at higher $p_T$ region. We can see also that the absolute size of the EW relative correction continuously grows up with the increment of $p_T$ at high $p_T$ region because of the Sudakov logarithms \cite{sudakov, fadin}. When $p_T^{l^+}$ goes up to $600~{\rm GeV}$, the relative EW correction can reach $-10\%$, while at the position around $p_T^H \simeq 600~{\rm GeV}$, the relative EW correction is about $-11\%$.

\par
\begin{figure*}[htbp]
\includegraphics[width=8cm]{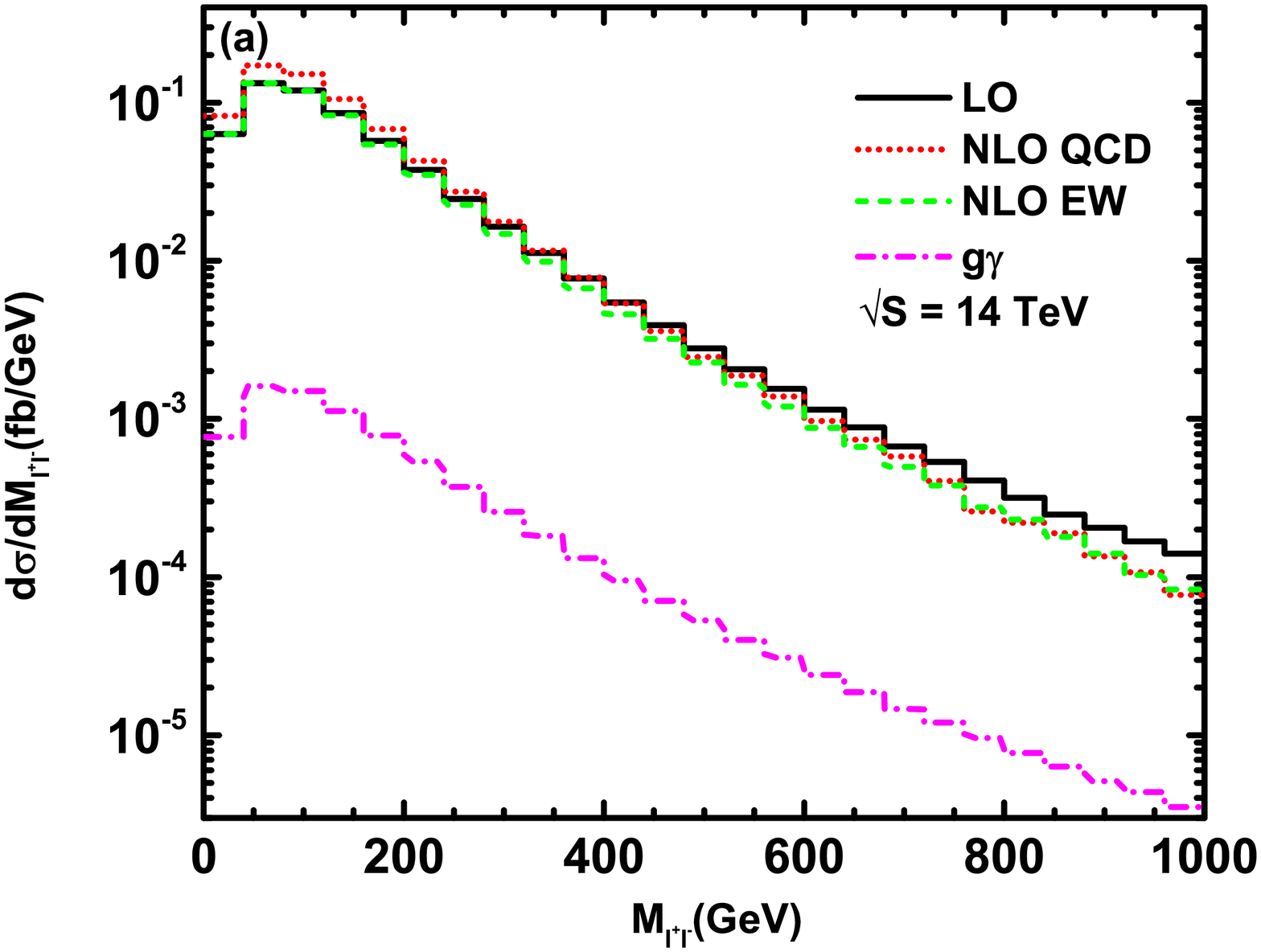}
\includegraphics[width=8cm]{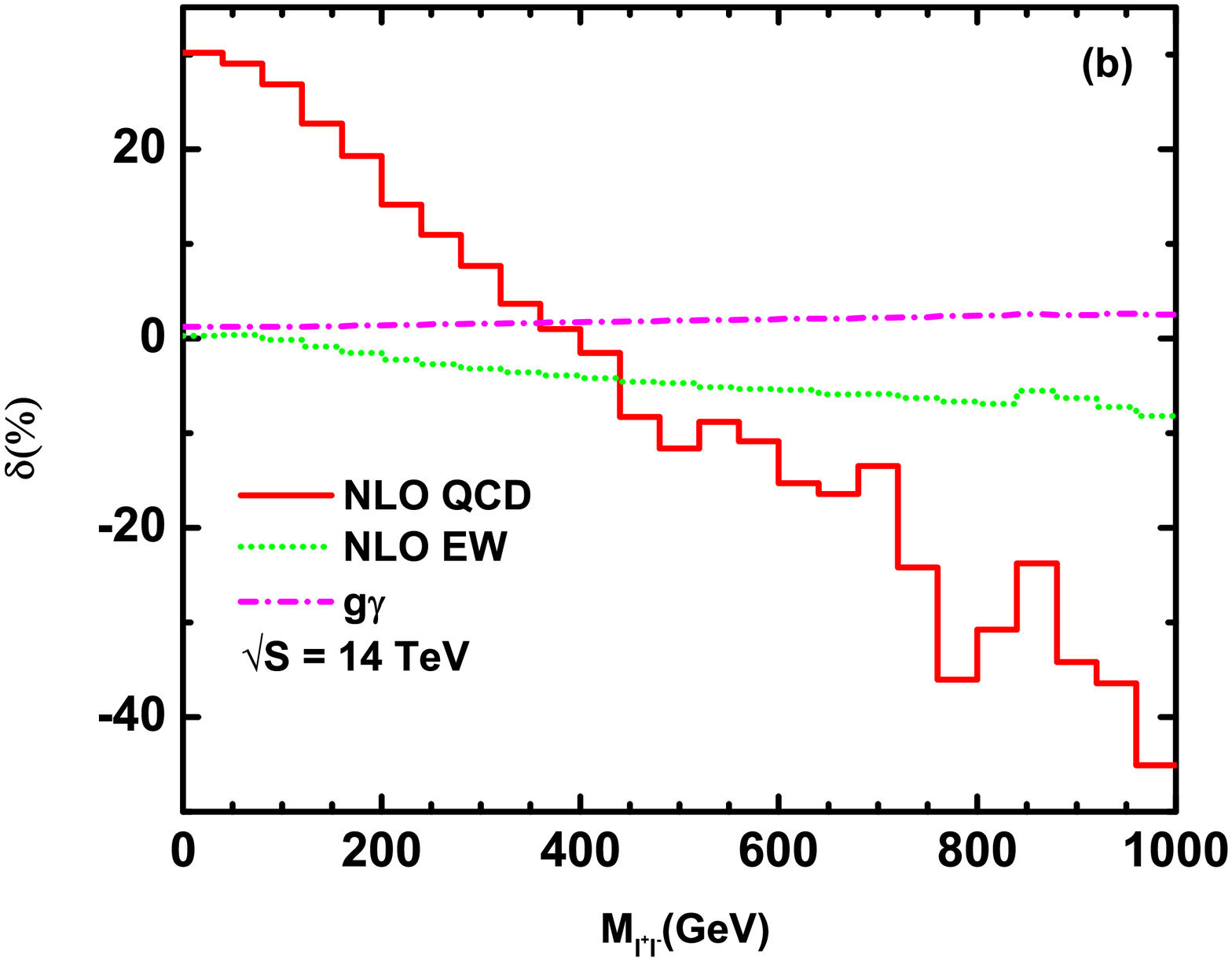}
\caption{
(a) The LO, NLO QCD, NLO EW corrected distributions of lepton pair invariant mass ($l^+l^- =e^+e^-$, $e^+\mu^-$, $\mu^+e^-$, $\mu^+\mu^-$) at the $14~{\rm TeV}$ LHC. The contributions from the $g\gamma \to t\bar{t}H$ subprocess are also shown there. (b) The corresponding relative corrections.  }
\label{Fig:mtt}
\end{figure*}
\begin{figure*}[htbp]
\includegraphics[width=8cm]{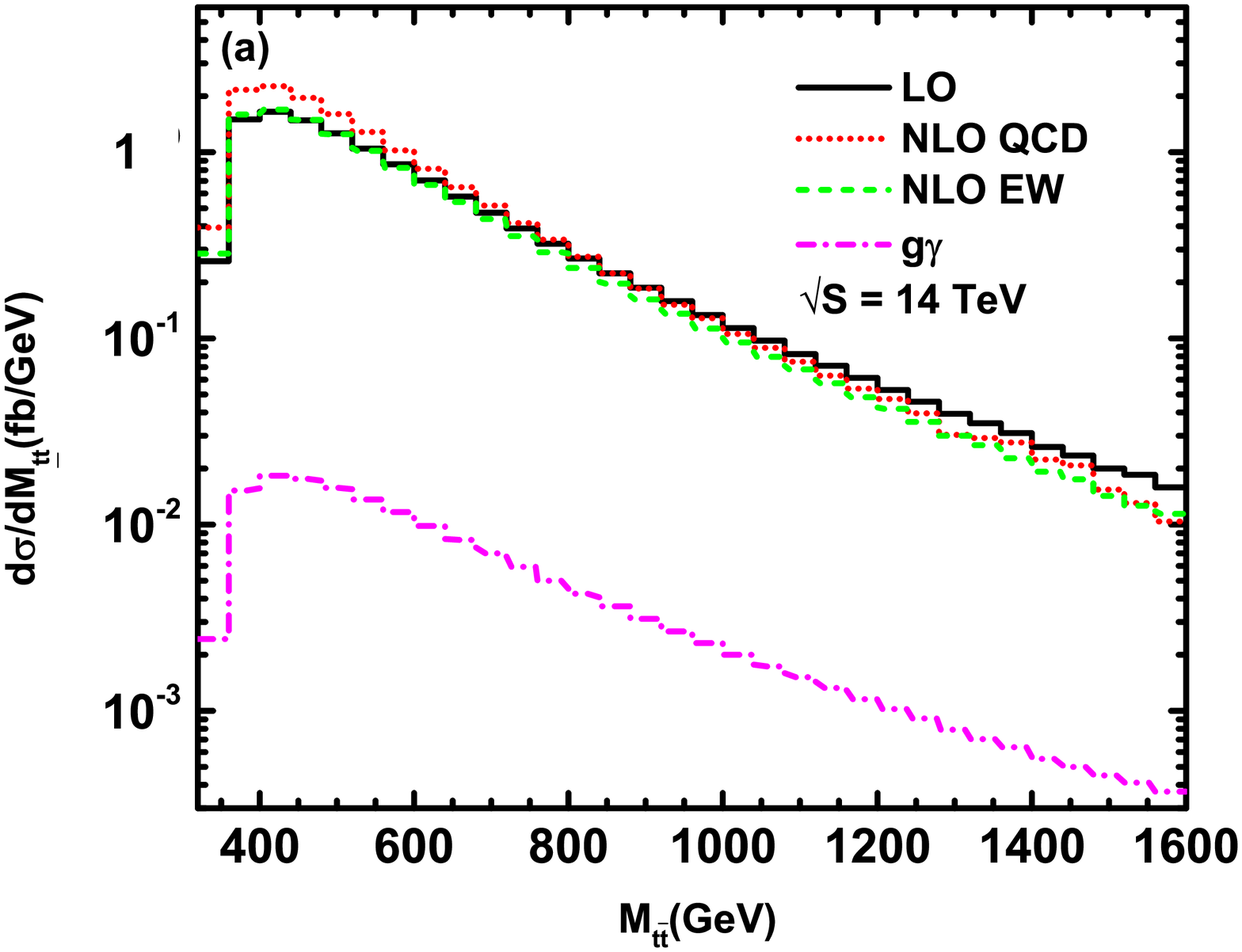}
\includegraphics[width=8cm]{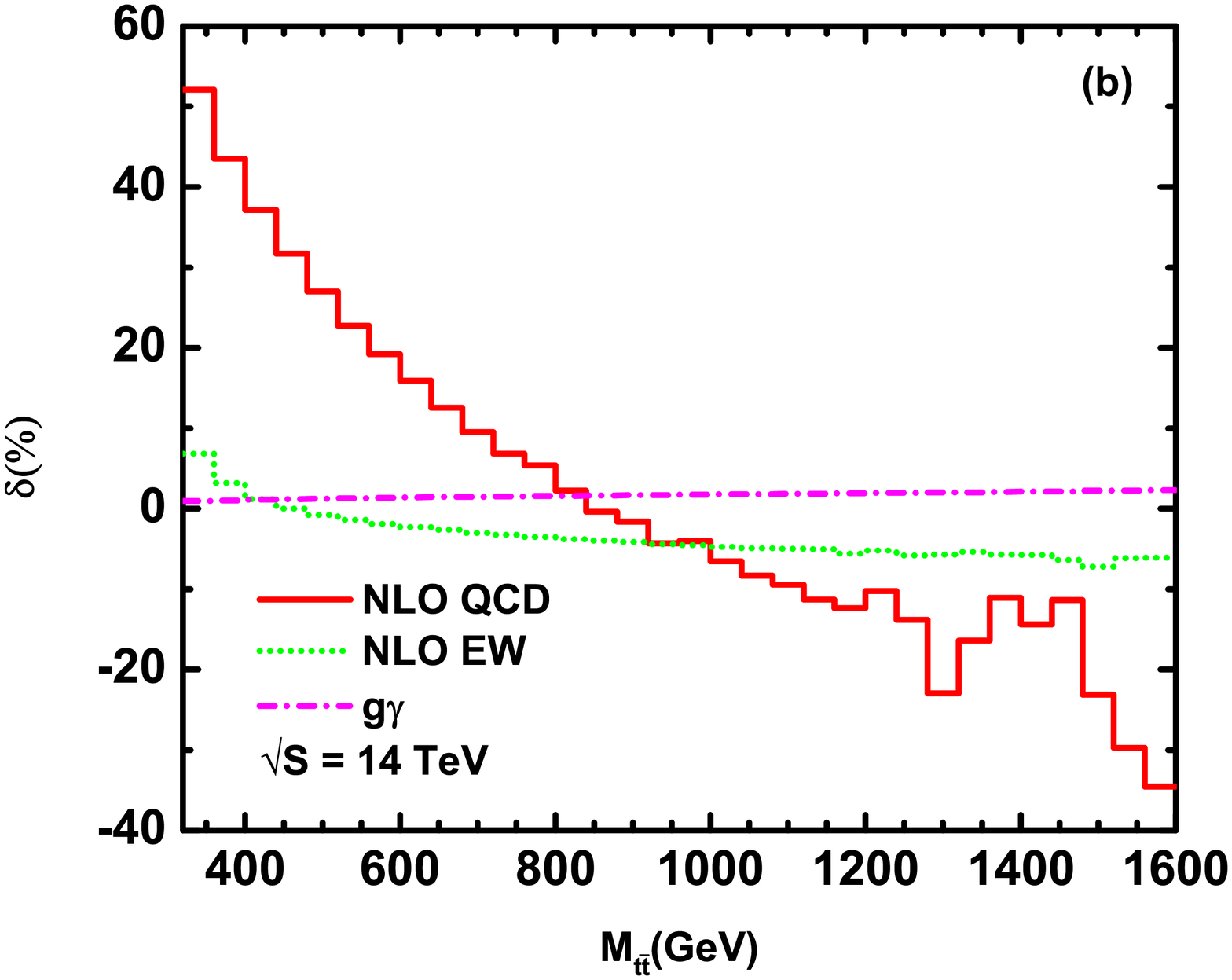}
\caption{
(a) The LO, NLO QCD, NLO EW corrected distributions of top pair invariant mass at the $14~{\rm TeV}$ LHC. The contributions from the $g\gamma \to t\bar{t}H$ subprocess are also shown there. (b)  The corresponding relative corrections.  }
\label{Fig:mtt-1}
\end{figure*}

In Fig.\ref{Fig:mtt}(a) and Fig.\ref{Fig:mtt-1}(a), we display the distributions of the invariant masses of the final lepton pairs $(M_{l^+l^-})$ and top quark pair $(M_{t \bar{t}})$ at the $14~{\rm TeV}$ LHC, respectively. The corresponding relative corrections are shown in Fig.\ref{Fig:mtt}(b) and Fig.\ref{Fig:mtt-1}(b) separately. We can see that the LO differential cross section $d\sigma_{LO}/dM_{l^+l^-}$ ($d\sigma_{LO}/dM_{t \bar t}$) is enhanced by the NLO QCD correction when $M_{l^+l^-}< 400~{\rm GeV}$ ($M_{t \bar t}< 840~{\rm GeV}$), and reduced if $M_{l^+l^-}$ ($M_{t \bar t}$) continues to become larger. The NLO EW corrected distributions demonstrate that the relative NLO EW corrections are always negative except in the region of $M_{l^+l^-}<80~{\rm GeV}$ and $M_{t \bar t}< 480~{\rm GeV}$. The relative NLO EW (QCD) correction amounts up to $-8\%$ ($-45\%$) for $M_{l^+l^-}=1000~{\rm GeV}$, and $-6\%$ ($-35\%$) for $M_{t \bar t}=1600~{\rm GeV}$.
One can see that the absolute relative NLO EW corrections are smaller than those of the NLO QCD corrections, but cannot be neglected.

\par
{\it Summary.-}
Precision predictions for $t \bar{t}H$ production at a hadron collider is very important in probing the top Yukawa coupling. In this letter, we present the calculations up to the QCD NLO and EW NLO for the process $pp\to t \bar t H$ with NWA top decays. We employ the most recent NNPDF2.3QED PDFs which include QED corrections and photon distribution function, and obtain the total NLO QCD and NLO EW correctied cross sections for the $pp \to t\bar{t}H$ process and the contributions from gluon-photon fusion subprocess at ${\cal O}(\alpha_s\alpha_{ew}^2)$. We provide the integrated cross sections at the $\sqrt{s}=14$, $33$ and $100~{\rm TeV}$ proton-proton colliders, and find that LO integrated cross sections are increased by the ${\cal O}(\alpha_s^3\alpha_{ew})$ correction while reduced by the ${\cal O}(\alpha_s^2\alpha_{ew}^2)$ correction. We give also the distributions for some important kinematic observables of the final particles after subsequent on-shell top decays ($t \to W b \to l \nu b$) at the $14~{\rm TeV}$ LHC. Due to the well-known EW Sudakov logarithms, the ${\cal O}(\alpha_s^2\alpha_{ew}^2)$ corrections become more and more sizable with the increments of the transverse momenta and invariant mass of final products. We conclude that besides the significant NLO QCD correction, the NLO EW correction is also worth being taken into account in precision measurement of the top Yukawa coupling at high energy hadron colliders.

\par
{\it Note Added.-}
After submission of this letter there appeared a paper on the same process \cite{Frixione}, where the authors provided the NLO effects from the weak and QCD corrections, whereas those of QED origin are ignored at $8$, $13$ and $100~ {\rm TeV}$ $pp$ colliders. In our paper we present all the effects from both NLO QCD and electroweak corrections. We find that our results are compatible with theirs apart from the QED corrections.

\par
{\it Acknowledgments.-}
This work was supported in part by the National Natural Science Foundation of China (Grants. No.11275190, No.11375008, No.11375171).

\end{document}